\documentclass[apjl]{emulateapj}
\usepackage{epsf}
\usepackage{amssymb,amsmath}
\usepackage{multirow}
\usepackage{rotating}
\usepackage{subfigure}
\usepackage{color}
\usepackage{epstopdf}

\newcommand\resetsubfigs{\setcounter{sub\@captype}{0}}

\def\aap{A\&A}
\def\apj{ApJ}

\def\apjl{ApJL}
\def\apjs{ApJS}
\def\mnras{MNRAS}

\def\pasj{PASJ}

\def\aaps{A\&A Supp.}

\def\apjs{ApJS}

\begin{document}
\title{Measuring A Truncated Disk in Aquila X-1}
\author{Ashley L. King\altaffilmark{1,2}, John A. Tomsick\altaffilmark{3}, Jon M. Miller\altaffilmark{4}, J\'er\^ome Chenevez\altaffilmark{5},  Didier Barret\altaffilmark{6,7}, Steven E. Boggs\altaffilmark{3},  Deepto Chakrabarty\altaffilmark{8}, Finn E. Christensen\altaffilmark{9}, William W. Craig\altaffilmark{3}, Felix F\"urst\altaffilmark{10}, Charles J. Hailey\altaffilmark{11}, Fiona A. Harrison\altaffilmark{10}, Michael L. Parker\altaffilmark{12}, Daniel Stern\altaffilmark{13}, Patrizia Romano\altaffilmark{14}, Dominic J. Walton\altaffilmark{10,13}, William W. Zhang\altaffilmark{15}   }
\slugcomment{Accepted to ApJL on February 22, 2016}
\altaffiltext{1}{KIPAC, Stanford University, 452 Lomita Mall, Stanford, CA 94305 USA, ashking@stanford.edu}
\altaffiltext{2}{Einstein Fellow}
\altaffiltext{3}{Space Science Laboratory, 7 Gauss Way, University of California, Berkeley, CA 94720-7450, USA}
\altaffiltext{4}{Department of Astronomy, University of Michigan, 1085 S. University Ave, Ann Arbor, MI 48109-1107, USA}
\altaffiltext{5}{DTU Space - National Space Institute, Technical University of Denmark, Elektrovej 327-328, 2800 Lyngby, Denmark}
\altaffiltext{6}{Universite de Toulouse, UPS-OMP, IRAP, Toulouse, France}
\altaffiltext{7}{CNRS, IRAP, 9 Av. colonel Roche, BP 44346, F-31028 Toulouse cedex 4, France}
\altaffiltext{8}{MIT Kavli Institute for Astrophysics and Space Research, Cambridge, MA 02139, USA}
\altaffiltext{9}{DTU Space, National Space Institute, Technical University of Denmark, Elektrovej 327, DK-2800 Lyngby, Denmark}
\altaffiltext{10}{Space Radiation Laboratory, California Institute of Technology, Pasadena, CA 91125 USA}
\altaffiltext{11}{Columbia Astrophysics Laboratory, Columbia University, New York, NY 10027, USA}
\altaffiltext{12}{Institute of Astronomy, Madingley Road, Cambridge, CB3 0HA, UK}
\altaffiltext{13}{Jet Propulsion Laboratory, California Institute of Technology, Pasadena, CA 91109 USA}
\altaffiltext{14}{INAF-IASF Palermo, Via Ugo La Malfa 153, 90146 Palermo, Italy}
\altaffiltext{15}{ NASA Goddard Space Flight Center, Greenbelt, MD 20771, USA}

\begin{abstract}
We present {\it NuSTAR} and {\it Swift} observations of the neutron star Aquila X-1 during the peak of its July 2014 outburst. The spectrum is soft with strong evidence for a broad Fe K$\alpha$ line. Modeled with a relativistically broadened reflection model, we find that the inner disk is truncated with an inner radius of $15\pm3 R_G$.  The disk is likely truncated by either the boundary layer and/or a magnetic field. Associating the truncated inner disk with pressure from a magnetic field gives an upper limit of $B<5\pm2\times10^{8}$~G. Although the radius is truncated far from the stellar surface, material is still reaching the neutron star surface as evidenced by the X-ray burst present in the {\it NuSTAR} observation.
\end{abstract}


\maketitle
\section{Introduction}
Neutron stars with magnetic fields on order of 10$^7$--10$^9$~G \citep{Mukherjee15} have the potential to truncate an accretion disk far from the stellar surface. The magnetic pressure exerted by the magnetic field lines can push out an accretion disk until the strength of the magnetic field is roughly equal to that of the ram pressure imposed by the accretion disk \citep[e.g.,][]{Pringle72,Illarionov75}. 

Though this model makes a clear prediction for the state of an accretion disk around a neutron star, direct confirmation is challenging. Disk truncation only occurs at a high enough magnetic field and low mass accretion rate, i.e. low pressure exerted from the disk. To be measurable, it requires detection techniques that are not dominated by uncertainties in the model. Continuum modeling, that relies on accurate determination of the luminosity of the disk to measure the inner radius, is dependent on mass, distance, and an unknown color correction factor \citep{Shimura95,Merloni00}. In general, the masses of neutron stars must lie in a small range \citep{Lattimer01}, but the distances and color correction factors are not well known, imposing large uncertainties on any radius measurements. 

A separate technique that relies on spectral fits to fluorescent Fe K$\alpha$ lines is free from these extrinsic uncertainties. The method assumes the line feature is reflected off the inner regions of the accretion disk, and precise determination of the red-wing of the Fe K$\alpha$ line gives the location of the inner edge \citep[see ][for a review in stellar-mass objects]{Miller06}. Several neutron star systems utilizing both {\it XMM-Newton} and {\it Suzaku} data show evidence for truncated disks \citep{Dai09,Cackett10,Papitto10,Papitto13,Egron13,DiSalvo15,Chiang15}. Unfortunately, these data sets can be biased by pile-up effects, as these sources are intrinsically bright, artificially narrowing the lines and biasing the inner disk radius toward higher \citep{Ballet99,Davis01,Miller10}. 

Though steps have been taken to either correct for pile-up effects \citep{Dai09,Papitto10,Cackett12,Papitto13,DiSalvo15,Chiang15} or observe sources at lower fluxes \citep{Cackett09,Miller11,Egron13,DiSalvo15}, using an instrument that does not suffer from pile-up is essential for examining many sources at the highest fluxes. This can unambiguously and robustly confirm previous measurements, and the model by which the magnetic field anchored to the neutron star can truncate the accretion disk at larger radii.

In order to accomplish such a task, we have employed {\it NuSTAR} \citep{Harrison13} to observe the iconic neutron star Aquila X-1 (Aql X-1). It is a low-mass X-ray binary with a K0 V companion \citep{Thorstensen78}. From observations of Eddington-limited photospheric radius expansion bursts its distance is 5.2$\pm0.7$ kpc \citep{Jonker04}, and is known to go through periodic outbursts \citep{Campana13} with both milli-second X-ray pulsations \citep{Zhang98,Casella08} and quasi-periodic oscillations \citep{Zhang98,Barret08}. During the decay of these outbursts, the luminosity drops steeply, suggesting that the source may go through a ``propeller stage''; a stage where material no longer accretes onto the neutron star but is thrown off by a strong magnetic field \citep{Asai13}. Estimates from modeling such behavior suggests the magnetic field strength is 0.6--31$\times10^8$~G \citep{Asai13,Campana14,Mukherjee15}. 

At this estimated magnetic field strength, Aql X-1 is a great candidate for measuring a truncated disk at higher mass accretion rates, i.e., higher luminosity states. In fact, Aql X-1 shows milli-second X-ray pulsations at frequencies of 550 Hz \citep{Zhang98,Casella08}, which corresponds to a Keplerian orbit of the inner disk radius at 11.5 $R_G$, where $R_G=GM/c^2$. An Fe K$\alpha$ line has been modeled with Gaussian components in previous observations, but then lacked both the sensitivity and resolution to analyze the line in detail \citep{Raichur11,Gungor14,Sakurai14}.

In this Letter, we present 32 ksec {\it NuSTAR} and 1.1 ksec {\it Swift} observations of Aql X-1 observations of Aql X-1 during an outburst in July 2014.

\section{Observations and Data Reduction}
\subsection{\it NuSTAR}
During an outburst in 2014 \citep{Meshcheryakov14}, we obtained a 32 ksec {\it NuSTAR} observation on UT 2014 July 17 (MJD 56855). The OBSIDs are 80001034002 and 80001034003, with exposure times of 11 and 21 ksec, respectively. The data were reduced with the standard {\tt NuSTARDAS}, v1.4.3, and {\tt CALDB} (20150312). Extraction regions for the OBSID  80001034002 had radii of 100 arcseconds for both the source and background regions. However, in OBSID 80001034003 was restricted to a source region with a radius of 50 arc seconds to avoid a chip gap. The background region was still taken with a 100 arcsecond radius. 

The spectra and responses from each focal plane module (FPM) were co-added with {\tt addascaspec} and {\tt addrmf}, similar to the analysis of Serpens X-1 \citep{Miller13} and 4U 1608-52 \citep{Degenaar15}. The cross-normalization constant between the initial FPMA and FPMB spectral fits differed by less than a percent. We then co-added the OBSIDs together, and binned the spectrum with a minimum signal to noise of 10, after background subtraction.  The total, co-added spectrum was then fit between 3--28 keV. Above 28 keV, the background dominates. The background subtracted spectrum had an average count rate of 120 counts s$^{-1}$ per FPM, resulting in a total of 7.7$\times10^6$ counts in the co-added spectrum.

\subsection{\it Swift}
We reduced the 1.1 ksec {\it Swift/XRT} observation taken on UT 2014 July 17 (MJD 56855) with the {\tt xrtpipeline} {\tt FTOOLS} software, version 6.17. This observation was strictly simultaneous with {\it NuSTAR} OBSID 80001034002 and taken in the Windowed Timing mode. 

In order to avoid the piled-up center, the {\it Swift} extraction region was an annulus centered on the source with inner radius of 3 pixels (7 arcseconds) and outer radius of 20 pixels (47 arcseconds) \citep{Romano06}. We also used a background region with a radius of 30 pixels (71 arcseconds). Exposure maps and ancillary files were created with {\tt xrtexpomap} and {\tt xrtmkarf}, respectively. The data was then grouped with a minimum of 20 counts per bin and fit between 0.7--10 keV. Below 0.7 keV, the windowed timing mode is known to have calibration artifacts\footnote{http://www.swift.ac.uk/analysis/xrt/digest\_cal.php}. The {\it Swift} background-subtracted spectrum has a total of 6.6$\times10^4$ counts with an average count rate of 60 counts s$^{-1}$.

\section{Analysis}
We fit the {\it NuSTAR} and {\it Swift} spectra with an absorbed, thermal Comptonization model, {\tt nthcomp} \citep{Zycki99} in {\tt XSPEC}, version 12.9.0. The absorption component, {\tt tbabs} \citep[set to wilm abundance;][]{Wilms00}, was allowed to vary above the measured line-of-sight Hydrogen column density $N_H=2.9\times10^{21}$ cm$^{-2}$ \citep{Kalberla05}. This accounts for both the line-of-sight column density as well as any intrinsic absorption near the source. 

\begin{figure*}
\includegraphics[width=.45\linewidth,angle=270,clip=true,trim=0cm 0cm -2cm -2cm]{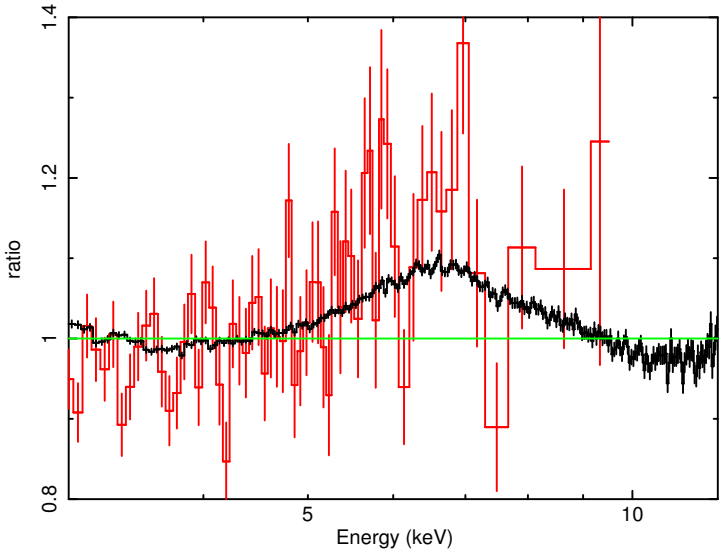}
\caption{This plots shows the {\it Swift} (red) and {\it NuSTAR} (black) residuals in the Fe K region after ignoring the regions between 5--10 keV and fitting with the {\tt nthcomp} model. A broad, almost symmetric line feature is strongly detected. The spectra have been binned by an additional factor of 20 for visual clarity. \label{fig:fekresids}}
\end{figure*}

The thermal Comptonization model has a power-law component with both a low and high energy cutoff. The low energy cutoff is set by the temperature of the seed photons. We assume this seed spectrum is a black body spectrum emitted from the boundary layer of the neutron star. The high energy cutoff is set by the electron temperature. In our fits, we allow both of these temperatures and the power-law index to vary. We find that the best fit model is quite poor with a  $\chi^2/\nu=7291.9/1202$. See Table 1. There are prominent residuals at 6.4 keV (Figure \ref{fig:fekresids}), indicating the presence of a reflection component in the Fe K$\alpha$ region.

  We add a Gaussian line to model the excess at $\sim$6.4 keV, restricting the energy range from 4--9 keV. Above and below this regime we expect contamination from the Compton hump and blurred reflection from other elements, respectively. Though Aql X-1 is one of the most well studied neutron stars, a broad Fe emission line at 6.4 keV has only been suggested \citep{Raichur11,Gungor14,Sakurai14}. However, it is strongly detected in our analysis, and is statistically required with $\Delta\chi^2=512$ for 3 additional degrees of freedom. The line is centered at 6.38$^{+0.05}_{-0.07}$keV, quite broad, with a full-width half-maximum of $1.98^{+0.07}_{-0.08}$ keV, and an equivalent width of 0.10$\pm0.04$ keV. 

\subsection{Reflection Model}
We next replace the Gaussian component with a more physical model, {\tt reflionx} \citep{Ross05}, which is convolved with a relativistic convolution kernel, {\tt relconv} \citep{Dauser10}. The {\tt reflionx} model is a version\footnote{https://www-xray.ast.cam.ac.uk/$\sim$mlparker/reflionx$\_$models/ reflionx$\_$alking.mod} of the \cite{Ross05} model that assumes a black body input spectrum illuminates an accretion disk producing fluorescence lines that are relativistically blurred, as well as a Compton hump.

\begin{figure*}
\centering
 \includegraphics[width=.75\linewidth,angle=270,clip=true,trim=0 0cm -5cm -2cm]{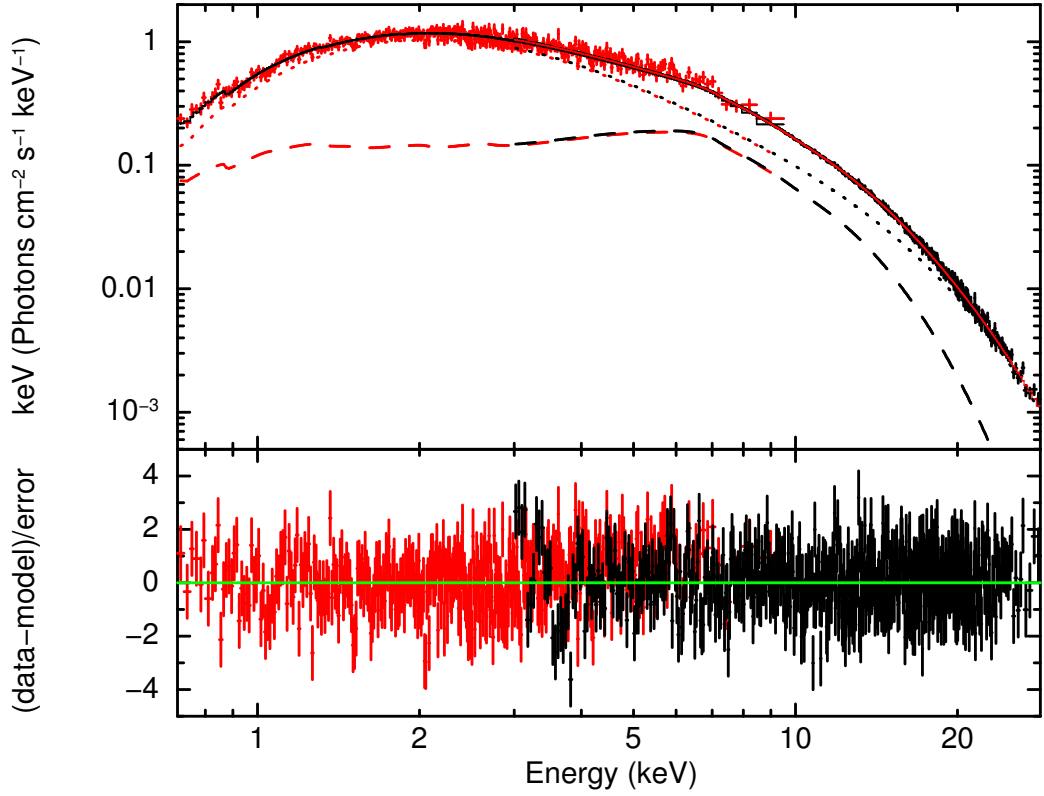}
\caption{The unfolded {\it Swift} (red) and {\it NuSTAR} (black) data in the top panel with the $\delta\chi =$ (data-model)/error in the bottom panel. The solid lines correspond to the co-added spectrum from the full model, while the dashed lines correspond to the reflection components. while the dotted line corresponds to the thermal Comptonization component ({\tt nthcomp}). \label{fig:nthcomp}}
\end{figure*}

We fix the spin parameter to $a=0.259$, which is equal to the spin frequency observed when Aql X-1  at a lower flux state \citep{Braje00,Casella08}, set the emissivity to $q=3$, and fit over the entire energy bands (0.7--28 keV). We note that when we relax the emissivity, the value is consistent with $q=3$, which is in agreement with a Newtonian geometry far from the neutron star, and also consistent with the radial extent of the disk in Section 4. The best fit parameters are shown in Table 1 and the model has a reduced chi-square of $\chi^2/\nu=1281.4/1196=1.07$ (Figure \ref{fig:nthcomp}). Fitting with just the {\it NuSTAR} spectra also results in a similar $\chi^2/\nu=639.8/610=1.05$, with little change in the fit parameters, indicating a reliable fit that is not driven by the {\it Swift} data alone. 

The resulting parameters from the Comptoniztation component  have a moderately steep spectral index of $\Gamma=2.59^{+0.11}_{-0.07}$, a low seed black body temperature of $T_{BB}=0.53\pm0.01$ keV, and an electron temperature of $kT_e=3.2^{+0.2}_{-0.1}$ keV. Though the black body and electron temperatures are quite low, the latter is still consistent with other observations of neutron stars, including Aql X-1 observations by {\it Suzaku} \citep{Raichur11}. 

The reflection component has a high ionization of $\xi=1900^{+500}_{-200}$ ergs cm s$^{-1}$, and an inclination of  $\theta={20^{+4}_{-3}}^\circ$. This inclination is consistent with infrared photometry measurements, $\theta<31^\circ$ \citep{Garcia99}.  The seed spectrum has a black body temperature of 1.95$^{+0.03}_{-0.04}$~keV. This temperature is likely approximating the continuum emission we observe, as the temperature is between the thermal disk and electron temperatures inferred from the {\tt nthcomp} component.

Interestingly, a large radius is required, $R_{in}=15\pm3~R_G$, indicating the disk is truncated at a large distance (Figure \ref{fig:radius}). The broadness of the line requires it to be located deep within the potential well where the orbital velocities are mildly relativistic, while the symmetry of the line requires it to be located far enough away as to not suffer severe relativistic Doppler beaming, apparent as a skewed red-wing, as seen in Serpens X-1, for example \citep{Miller13}. 

As the reflection model is dependent on the underlying continuum model, we also fit the data with a separate continuum model  assuming thermal Comptonization. This model includes a thermal black body to model the neutron star boundary layer, a thermal disk black body to model the accretion disk, and a phenomenological power-law. The fit is much worse; $\chi^2/\nu= 1594.6/1177=1.35 $, and has large residuals in the Fe K region (5--7 keV) where the black body and disk black body overlapped. However, a large radius is still required $R_{in}=20\pm3~R_G$, indicating the robustness of our truncated disk measurement.

\subsection{Type I X-ray Burst}

\begin{figure*}
\subfigure[\label{fig:1}][]{\includegraphics[width=.45\linewidth,angle=0,angle=0,clip=true,trim=0 -.5cm 0cm 1cm ]{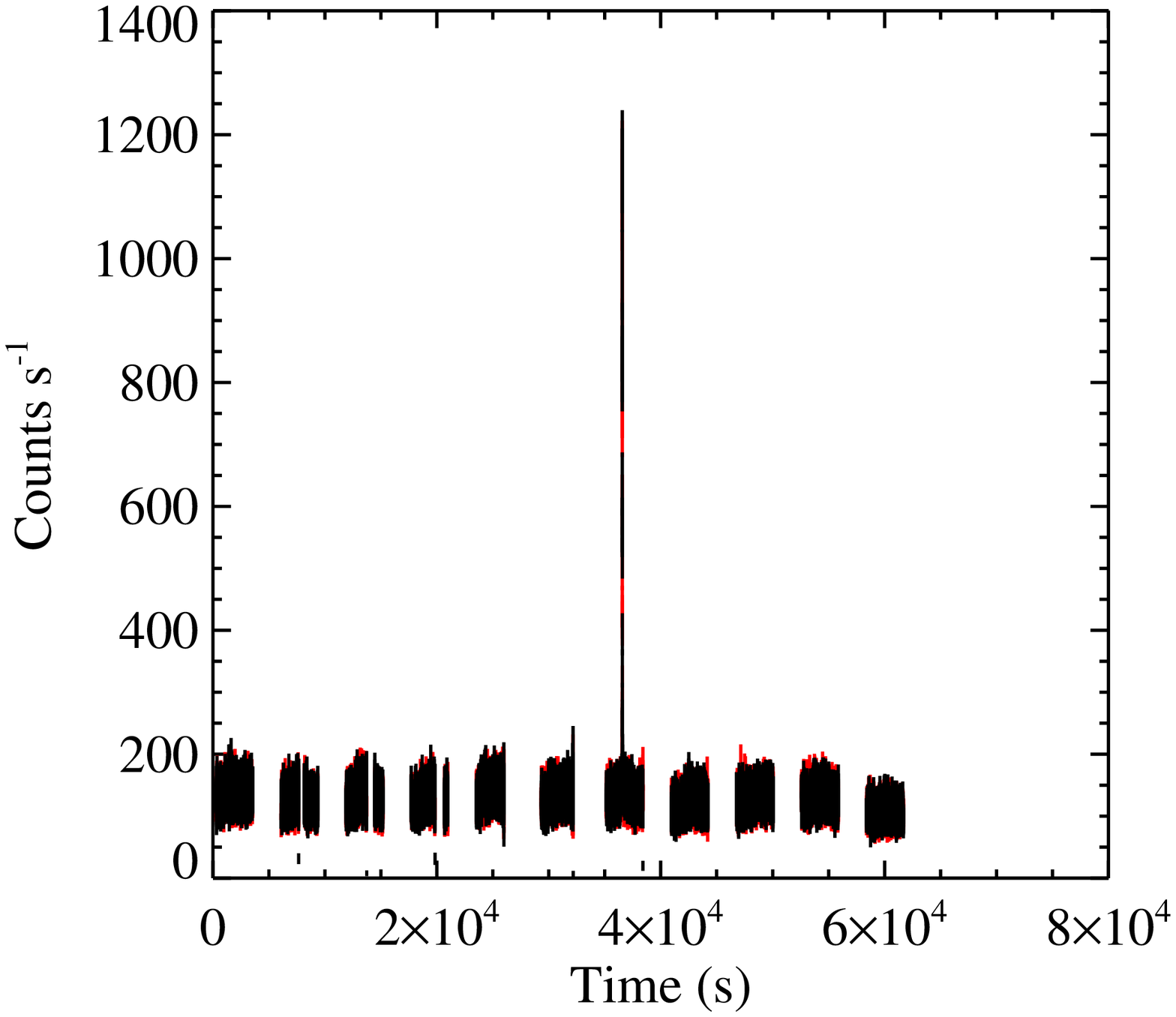}}
\subfigure[\label{fig:1}][]{\includegraphics[width=.45\linewidth,angle=0,angle=0,clip=true,trim=0 -.5cm 0cm 1cm ]{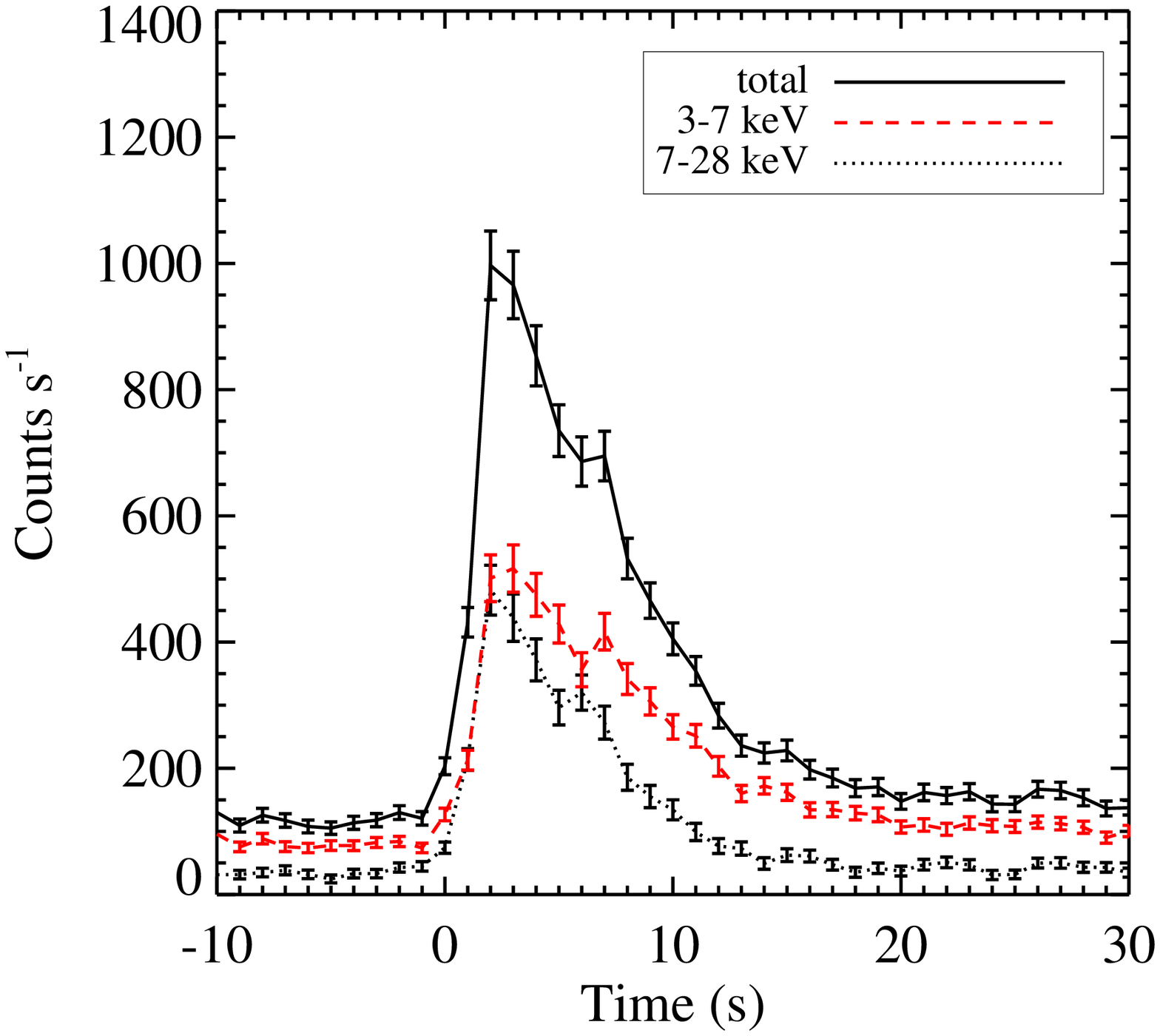}}

\subfigure[\label{fig:1}][]{\includegraphics[width=.45\linewidth,angle=0,clip=true,trim=0cm -.5cm 0 1cm]{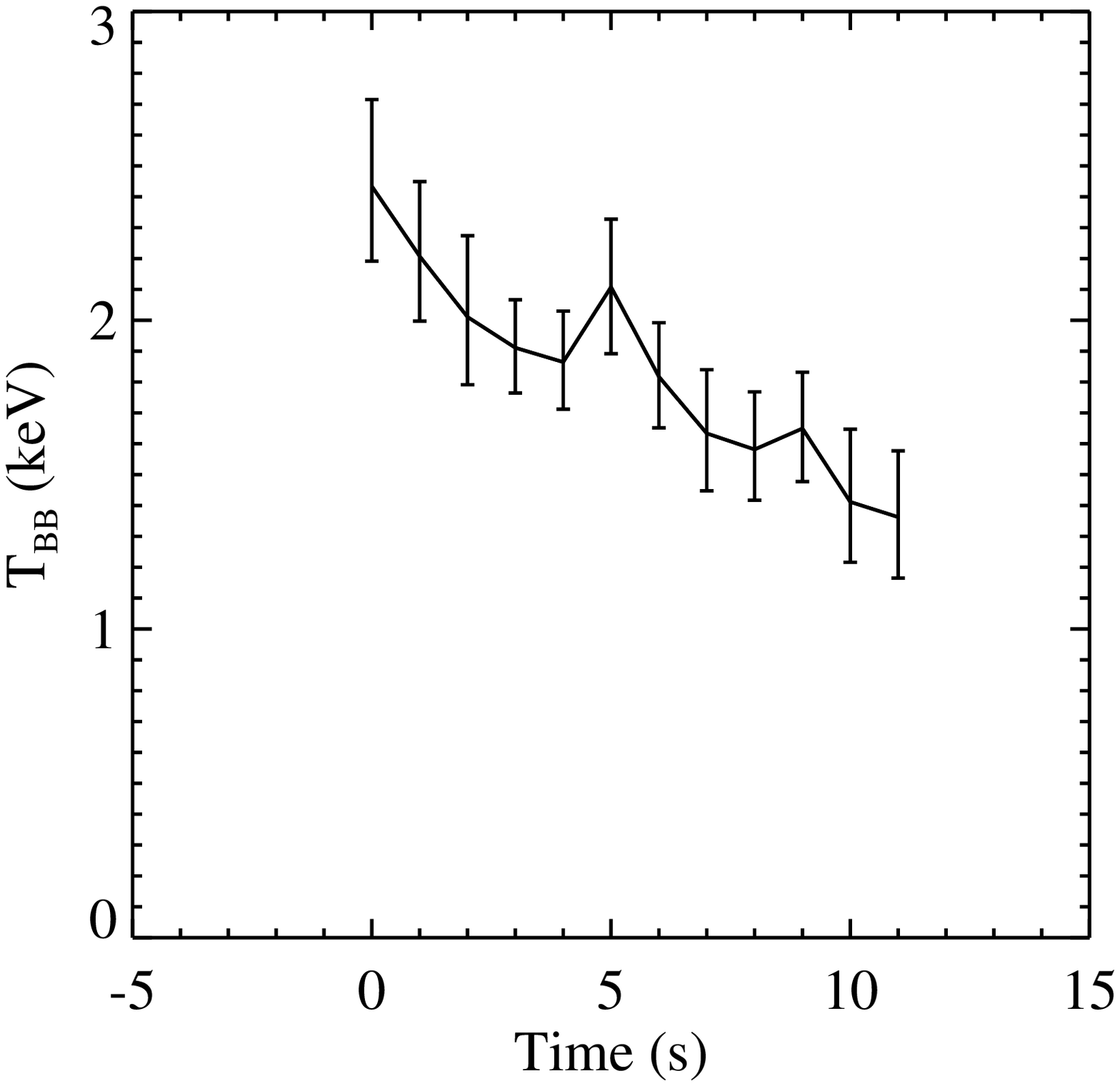}}
\subfigure[\label{fig:1}][]{ \includegraphics[width=.45\linewidth,angle=0,angle=0,clip=true,trim=0 -.5cm 0cm 1cm]{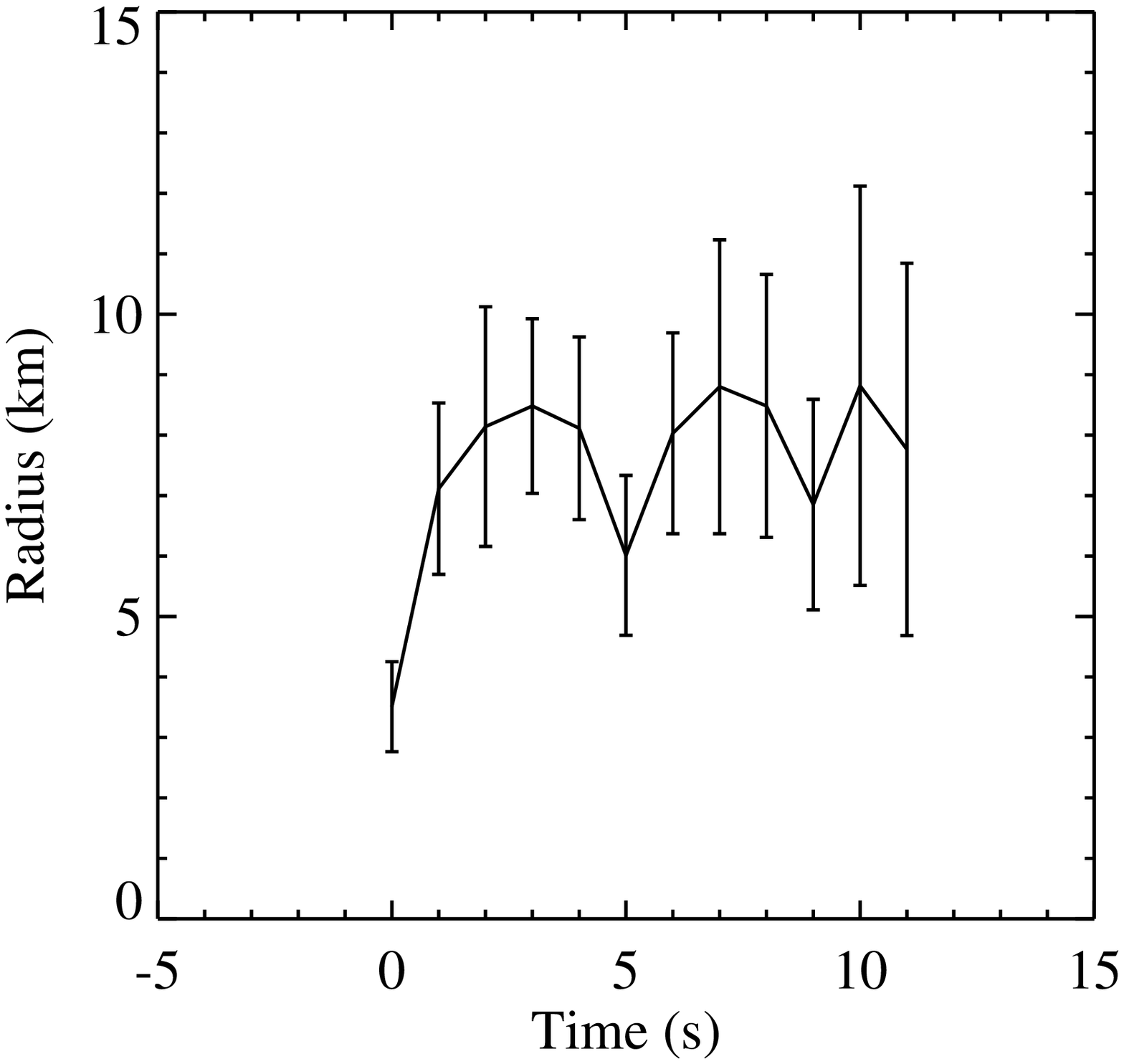}}
\caption{Panel (a) shows the FPM A (black) \& B (red)  1s 3--28 keV binned light curve from the {\it NuSTAR} OBSID 80001034003, with a clear X-ray burst, shown as the co-added spectrum (solid black) in detail in panel (b). The hard band (7--28 keV, black dotted) decays faster than the soft band (3--7 keV, red dashed), indicating a cooling tail during the burst. Panel (c) shows the measured black body temperature and its decay, and panel (d) shows the inferred black body radius during this X-ray burst, which expands from $3.5\pm0.7$ km to a roughly constant $\sim 8$ km.  \label{fig:burst} }
\end{figure*}

\begin{figure*}
\includegraphics[width=.6\linewidth]{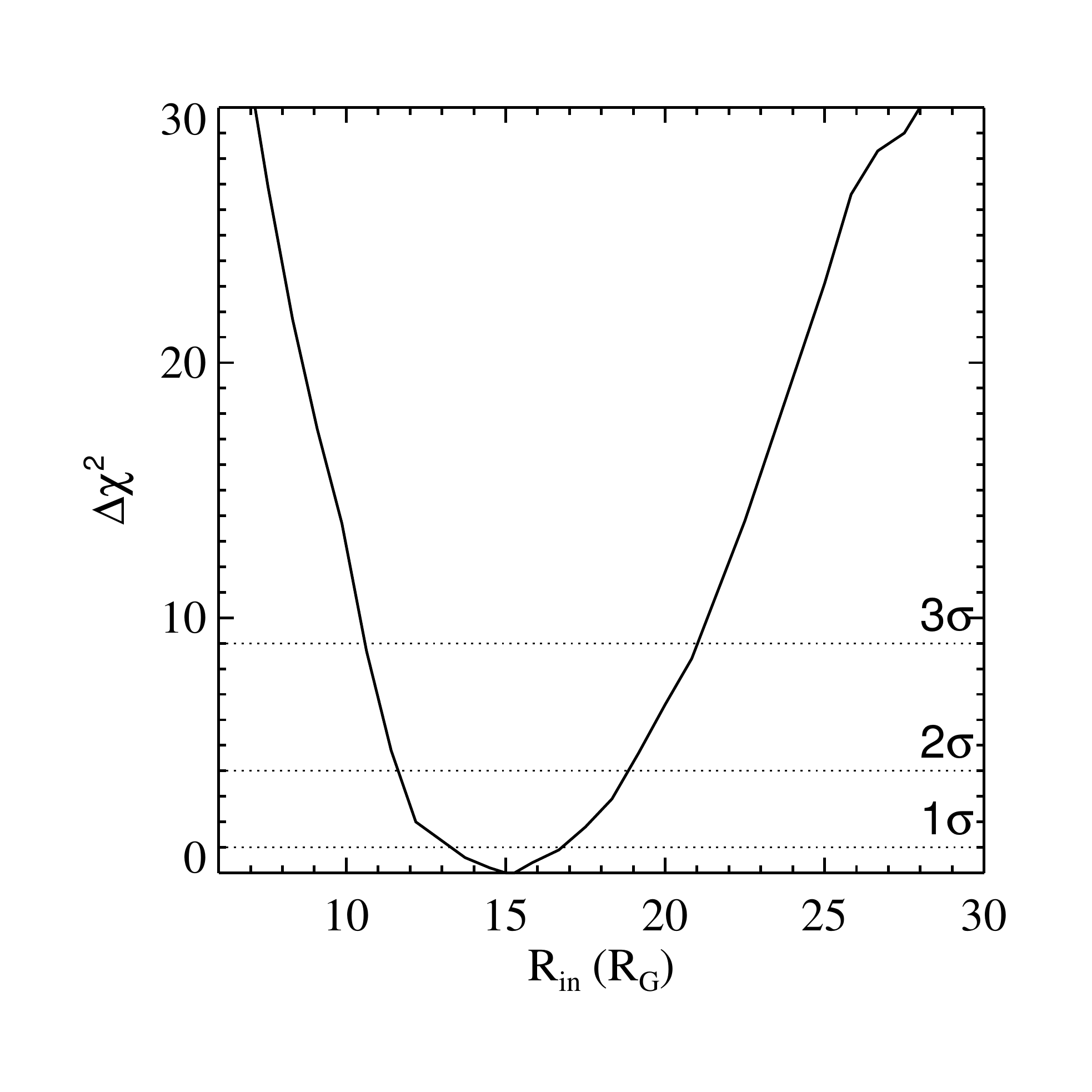}
\caption{Inner radius delta chi-squared distribution in the relativistic reflection. The inner radius is not consistent with the inner most stable orbit of 5.2$R_G$ or the neutron star surface ($<14$ km or 6.3$R_G$ for 1.5 M$_\odot$), indicating that it is truncated far from the neutron star surface. Associating this with the magnetic field, we find the inferred strength to be $5\pm2\times10^8G$. \label{fig:radius} }
\end{figure*}

During the {\it NuSTAR} observation, an X-ray burst occurred. Figures \ref{fig:burst}a \& \ref{fig:burst}b  show the light curve surrounding the burst, which lasted less than 20s. X-ray bursts, specifically Type I bursts, commonly result from unstable thermonuclear H/He burning in the surface layers of the neutron star after a critical mass has been accreted onto the surface \citep[see ][for a review]{Parikh13}. During some bursts, the radiation pressure may reach the Eddington limit so the burning layer expands, lifting off from the surface, thus leading to the expansion of the photosphere of the neutron star.

The 3--28 keV {\it NuSTAR} light curve of the burst shows a duration of $\sim$13~s above 25\% of the peak intensity  \citep[see ][]{Galloway08}, with a rise of 2s and an exponential decay timescale of 5.5$\pm0.2$s (Figure \ref{fig:burst}a \& \ref{fig:burst}b). The decay is shorter in the upper end (7--28 keV) of the energy band than in the lower end (3--7 keV), indicative of cooling during the burst tail (Figure \ref{fig:burst}b). We perform a time-resolved spectral analysis of the burst by holding constant the persistent emission obtained from the 1400~s prior to the burst, and modeling the burst emission with an absorbed black body, {\tt bbodyrad}. We binned the spectra into 1~s intervals containing a minimum of 100 counts in the co-added FPMA and FPMB spectra. 

The burst temperature peaks at the start of the burst, cooling throughout its duration as seen in Figure \ref{fig:burst}c. The burst emission is marginally consistent with a photospheric radius expansion lasting less than 1s, suggesting that the burst may not have reached its Eddington limit, $L_{Edd}$ (Figure \ref{fig:burst}d). At the source distance of 5.2$\pm0.7$ kpc \citep{Jonker04}, we measure the unabsorbed bolometric (0.1--100 keV) peak luminosity = $2.0^{+1.2}_{-0.8}\times10^{38} $ erg/s. This is 1.2~$L_{Edd}$, assuming a solar composition, 1.5~$L_{Edd}$ assuming pure Hydrogen, and 0.7~$L_{Edd}$ assuming a pure Helium atmosphere.

We conclude from our analysis that the burst observed by {\it NuSTAR} was a thermonuclear flash due to the unstable burning of a He-rich layer on the neutron star surface. Detection of this burst indicates material is still reaching the surface of Aql X-1, even when the disk is truncated at such a large distance.

\section{Discussion}

We have observed Aql X-1 in the July 2014 outburst. The data require a significant reflection component, characterized by the Fe K$\alpha$ broad emission line. If we fit this feature with a physical reflection model, we find that the line has a relatively low ionization and viewing angle, as well as a large inner radius ($R_{in}$=$15\pm3$ $R_G$). 

The radius we measure is formally ($3\sigma$) inconsistent with both the innermost stable circular orbit for this rotating neutron star \citep[$R_{ISCO}=5.2R_G$,][]{Bardeen72} and the neutron star surface ($R_{surface}$=10--14km = 4.5--6.3$R_G$, depending on the equation of state \citep[][ Figure \ref{fig:radius}]{Lattimer01}. This means that the radius is truncated away from the neutron star surface. 

This truncation is likely the result of either a state transition associated with a receding disk \citep{Esin97}, a boundary layer \citep{Popham01,Dai10} or a magnetic field exerting a pressure on the disk \citep{Illarionov75}. A receding disk from a state transition is typically associated with low-luminosity and X-ray spectra that are dominated by a hard power-law \citep{Esin97}. The Aql X-1 spectra presented here are taken during the middle of the outburst, in a soft, high luminosity state ($2\times10^{37}$ ergs s$^{-1}$). Thus, a state transition appears to be unlikely, especially as such a state transition generally occurs at few$\times10^{36}$ ergs s$^{-1}$ in Aql X-1 \citep{Asai13}. 

A boundary layer is a viable mechanism for truncating the disk. \cite{Popham01} discuss Newtonian models for the boundary layer, varying mass accretion rate, viscosity and spin frequency of the neutron star. We estimate the mass accretion rate during this observation to be $5.2\times10^{-9}$ M$_\odot$yr$^{-1}$, which is extrapolated from the 0.1-100 keV luminosity and assuming a radiative efficiency of $\eta=0.067$, i.e., the efficiency given the binding energy at 15 $R_G$. At this mass accretion rate, \cite{Popham01} estimate the boundary layer to extend to $R_{B}\sim 7.8 R_G$. However, changes in viscosity and rotation could increase this to be consistent. 

Alternatively, the boundary layer may actually be smaller than the disk truncation radius, and the magnetic field would be responsible for the truncation. We use the following relation given by \cite{Illarionov75} to derive the magnetic field, 
\begin{equation}
R_{in}=4\times10^8~B_{11}^{4/7}\dot{m}_{15}^{-2/7} M^{-1/7}~{\rm cm}
\end{equation}
where $B_{11}$ is the magnetic field in units of $10^{11}$~G, $\dot{m}_{15}$ is the mass accretion rate in units of $10^{15}$~g~s$^{-1}$, and $M$ is the mass of the neutron star in solar masses. Utilizing the measured inner radius from the reflection models, and a mass of 1.5$M_\odot$, we find a magnetic field strength of $B=5\pm2\times10^8G$. This is consistent with previous estimates for Aql X-1 of B=0.6--30$\times10^{8}$~G \citep{DiSalvo03,Asai13,Campana14,Mukherjee15}. However, we note that our value of B is an upper limit, as the boundary layer could still be truncating the disk.

Associating the truncation radius ($R_{in}=15\pm3$ $R_G$) with a Keplerian frequency, we find that it corresponds to a frequency of $\nu=370\pm110$ Hz. This is slightly lower then the spin frequency of 550.27 Hz \citep{Casella08}, which would correspond to an inner radius of $R_{in}=11.5R_G$, assuming 1.5M$_\odot$.  However, our measurements of the inner radius are still within $3\sigma$ of 11.5$R_G$, i.e., the co-rotation radius, and a slightly smaller mass would bring these measurements into better agreement (1$\sigma$). Furthermore, as we observe a Type I X-ray burst, this indicates that Aql X-1 is still accreting during the {\it NuSTAR} observation. When assuming the magnetic field is truncating the disk, it implies that the magnetosphere does not extend past the co-rotation radius, which would limit the accretion onto the neutron star in a ``propeller stage''. Such a stage is observed, though at lower luminosities ($\sim5\times10^{36}$ ergs s$^{-1}$) in Aql X-1 \citep{Campana14}.

Though evidence for accretion is observed via the Type I burst, we do not find evidence for X-ray pulsations. Thus, the accreting material may be channeled along the field lines to the polar caps \citep{DeLuca05}, though this requires a large ``hot spot'' or the emission region to align with the spin axis in order to be consistent with the lack of pulsations \citep{Lamb09}. 

\section{Summary}
{\it NuSTAR} and {\it Swift} caught Aql X-1 during the bright outburst in July 2014. Residuals from a simple, thermal Comptonization model still show broad features in the Fe K$\alpha$ region. Though this feature is almost symmetric, as indicated by relatively good fits with a simple Gaussian model, the breadth suggests emission deep in the neutron star potential well. A relativistic reflection model indicates the truncation radius is $R_{in}=15\pm3 R_G$. Associating the truncation radius with a Keplerian frequency, we find it is consistent with the spin frequency \citep{Zhang98,Casella08}. Associating this truncation with the magnetic field, we infer an upper limit to the magnetic field strength of $5\pm2\times10^8$~G. In addition, we can infer material is still reaching the surface in this configuration since a type I X-ray burst is detected in the {\it NuSTAR} observation. 

\begin{acknowledgements}
The authors thank the referee for their invaluable comments. ALK would like to thank the support for this work, which was provided by NASA through Einstein Postdoctoral Fellowship grant number PF4-150125 awarded by the Chandra X-ray Center, operated by the Smithsonian Astrophysical Observatory for NASA under contract NAS8-03060. PR acknowledges financial contribution from contract ASI-INAF I/004/11/0 and ASI-INAF I/037/12/0. JC thanks financial support from ESA/PRODEX Nr. 90057. This work made use of data from the {\it NuSTAR} mission, a project led by the California Institute of Technology, managed by the Jet Propulsion Laboratory, and funded by the National Aeronautics and Space Administration. This research has made use of the {\it NuSTAR} Data Analysis Software (NuSTARDAS) jointly developed by the ASI Science Data Center (ASDC, Italy) and the California Institute of Technology (USA). 
\end{acknowledgements}

\begin{deluxetable*}{lllllll} 
\tabletypesize{\small}
\tablecolumns{7} 
\tablewidth{0pc} 
\tablecaption{Model Parameters} 
\tablehead{ 
\colhead{Model} & \colhead{parameter} & \colhead{\it NuSTAR} & \colhead{\it NuSTAR}  \\
 & &   \colhead{+ {\it Swift}} &  \colhead{+ {\it Swift}}\\
& & A & B  }
\startdata
 &  $N_H (10^{22}$ cm$^{-2}$) & 0.29$^{+0.01}_*$ & 0.43$\pm0.02$\\
   \\
  \multirow{4}{*}{ \rotatebox{90}{ nthComp}} &   $\Gamma_{\rm nthComp}$ & 2.10$\pm0.01$ & 2.59$^{+0.11}_{-0.07}$ \\
  & $kT_{e}$ (keV) & 2.30$\pm0.01$ & 3.2$^{+0.2}_{-0.1}$\\
   & $kT_{BB}$ (keV) & 0.60$\pm0.01$  & 0.53$\pm0.01$\\
   & $N_{\rm nthComp}$ &  0.72$^{+0.02}_{-0.01}$ & 0.93$\pm0.03$ \\
  \\
   \multirow{3}{*}{ \rotatebox{90}{relconv(reflionx)}} & $\theta$ ($^\circ$)  &   -  & 20$^{+4}_{-3}$ \\
  &  $R_{in} (R_G)$  &  - & 15$\pm3$\\
 & $\xi$ (ergs cm s$^{-1}$)    &  -  & 1900$^{+500}_{-200}$ \\
 & $Z_{Fe}$ &  - & 0.50$^{+0.13}_{*}$ \\
 & $kT$ (keV) &  - &  1.95$^{+0.03}_{-0.04}$\\
  & $N_{Refl}$   &  - & 1.10$^{+0.12}_{-0.06}$  \\
 \\
&  $C_{\rm XRT/NuSTAR}$ &  1.10$\pm0.01$ & 0.98$\pm0.01$\\
 \\
 & $\log(F_{0.1-100 keV})$ (ergs  s$^{-1}$ cm$^{-2}$) &    -(7.988$_{-0.002}^{+0.001}$)  & -(7.930$\pm0.003$) \\
 &  $\log(L_{Bol}$) (ergs s$^{-1}$) & 37.3$\pm0.1$ & 37.3$\pm0.1$ \\
&  $<$B (10$^8$~G) &  - &  5$\pm2$ \\
 \\
 &   $\chi^2/\nu_{\rm dof}$   &  7292/1201=6.07 & 1281.4/1196=1.07 \\
\enddata
\tablecomments{The list of fit parameters for thermal Comptonization model ({\tt nthcomp}, A) or a relativistic reflection model (B). A poor fit and residuals shown in Figure 1, indicate Fe K$\alpha$ feature. The reflection model gives a statistically acceptable fit, and  the disk is truncated at $15\pm3$ $R_G$. We assumed a distance of 5.2$\pm0.7$ kpc \citep{Jonker04} and a mass of 1.5 M$_\odot$ for calculating the luminosity and upper limit of the magnetic field strength. 
\
*parameters have hit the bounding limits. }
\end{deluxetable*}





\end{document}